# Lightweight Machine Learning for Classifying Disaster-Related Social Media Posts

Chengjun Jiang
School of Computer Science
University of Nottingham
psxcj11@nottingham.ac.uk

Milena Radenkovic
School of Computer Science
University of Nottingham
milena.radenkovic@nottingham.ac.uk

***Abstract*—Classifying disaster-related social-media posts into humanitarian information labels is challenging because individual messages are typically short and context-limited, humanitarian categories can overlap semantically, and the CrisisBench label distribution is strongly imbalanced. This study investigates whether lightweight lexical machine-learning methods can provide effective and computationally efficient classification under these conditions. Using the CrisisBench v1.0 English humanitarian task, comprising 87,455 records across predefined Train, Development and Test splits, we conduct a controlled comparison of TF-IDF representations and lightweight classifiers, with Macro-F1 as the primary model-selection metric. Beyond aggregate predictive performance, the study jointly examines per-label behaviour, directional error patterns, relative decision margins, computational requirements, and the effect of domain-specific training through a tropical-cyclone specialist model. The selected balanced LinearSVC with unigram–bigram TF-IDF and C=0.5 achieved an Accuracy of 0.8282, a Macro-F1 of 0.6578 and a Weighted-F1 of 0.8315 on the held-out Test split. Lower decision margins were substantially associated with prediction errors, while the tropical-cyclone specialist achieved modestly higher in-domain aggregate performance with substantially lower feature and fitting requirements. The study contributes a controlled, label-sensitive and resource-aware evaluation of lightweight humanitarian text classification that characterises not only aggregate performance, but also where these models fail and how their behaviour changes under domain-specific training.**

## I. Introduction

Social media has become a vital communication channel during disasters, enabling the public to report local conditions, share early warnings, request assistance, and describe the impact of disasters as events unfold. Research on two natural-hazard events found that posts from people near the affected areas contained information that could contribute to situational awareness [1]. However, the flow of crisis-related information also includes repetitive headlines, personal reactions, rumours, irrelevant discussions, and information of limited humanitarian value. Since posts are typically brief, informal, and context-dependent, much of this content is difficult to utilise directly without further processing. Consequently, information overload, information prioritisation, and brief, informal text represent core challenges for social media analysis in emergency situations [2].

This study investigates whether lightweight TF-IDF classifiers can provide effective and computationally efficient humanitarian classification under three closely related difficulties: limited textual context, semantic overlap between labels, and severe class imbalance. The investigation uses the CrisisBench v1.0 English humanitarian task [4], containing 87,455 records and 11 target labels. Only cleaned post text is provided as predictive input, requiring the classifiers to distinguish humanitarian meanings without relying on event or source metadata.

The classification problem is non-trivial because crisis posts are typically short, informal and context-limited, and may contain abbreviations, hashtags, mentions, URLs or incomplete sentences [2]. Humanitarian meanings can also overlap: a single post may describe infrastructure damage and urgent needs even though the benchmark requires one final label. In addition, not_humanitarian accounts for approximately 59% of the dataset, whereas some minority labels represent less than 1%. Macro-F1 is therefore used as the primary model-selection metric, with Accuracy, Weighted-F1 and per-label measures providing complementary evidence [23].

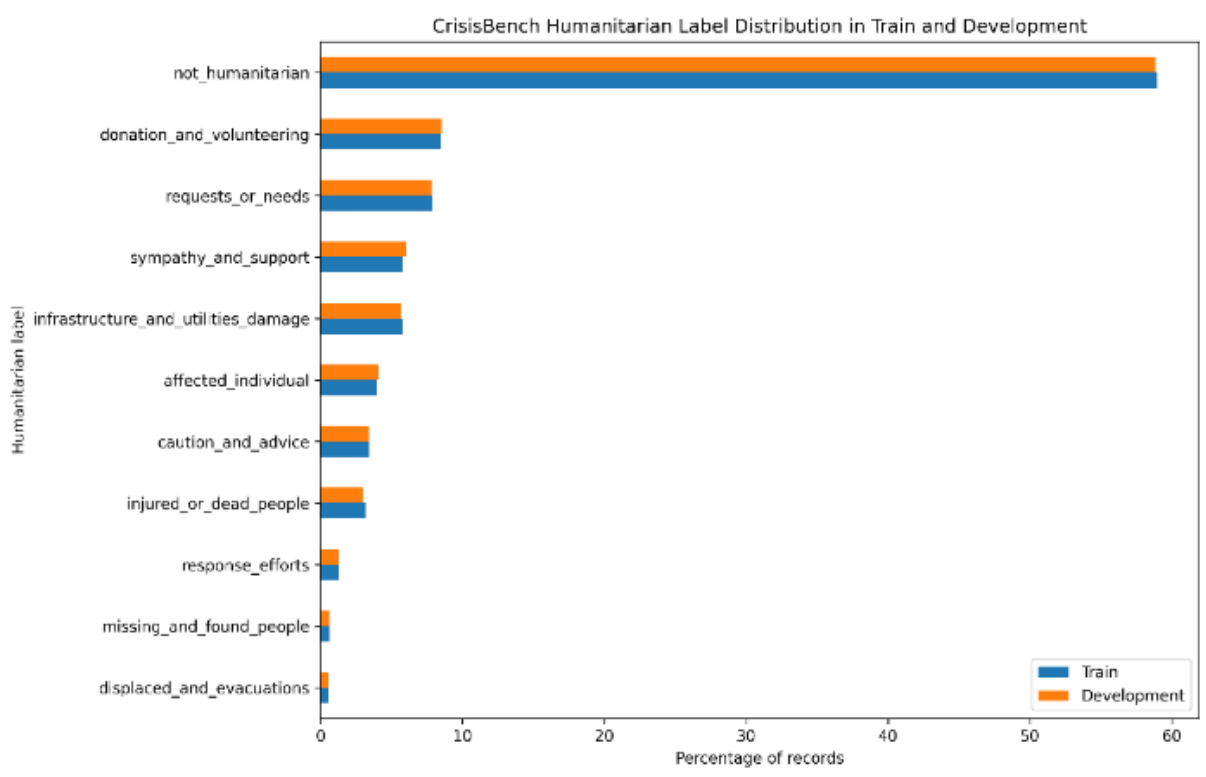

Figure 1: Percentage distribution of CrisisBench humanitarian labels in the Train and Development splits

The distinctive empirical contribution of this study is a controlled, label-sensitive and resource-aware investigation of lightweight humanitarian text classification. The evaluation extends beyond aggregate benchmark scores by combining classifier and representation comparison with per-label performance, directional error analysis, relative decision margins and measured computational requirements. It further examines how domain restriction changes this performance–resource trade-off by comparing a general multi-disaster classifier with a tropical-cyclone specialist. Together, these analyses characterise which lightweight configurations perform effectively, where humanitarian label distinctions remain difficult, and how predictive behaviour and computational requirements change under domain-specific training.

## II. Related Work

### *A. Crisis Information Classification and Datasets*

AIDR demonstrates how machine learning and human annotation can be combined to classify crisis-related messages into user-defined categories, such as needs and damage [3].

A manually annotated corpus covering nineteen crisis events utilised a humanitarian classification system based on the classification scheme used by the United Nations Office for the Coordination of Humanitarian Affairs [5]. The corpus includes information on casualties, missing or accounted-for persons, displacement, damage to infrastructure, donations, early warnings, support information, and other relevant or irrelevant information. The final labels were determined collaboratively by multiple annotators [5].

As the events and research objectives vary across projects, the labelling schemes for different datasets also differ. CrisisBench addresses this inconsistency by mapping semantically related source labels to a shared scheme and excluding records that cannot be consistently converted to the target task [4]. Whilst this improves comparability between models, the harmonised scheme does not preserve every distinction present in the source datasets, since semantically related labels are mapped to shared categories and some source-specific labels or records are excluded during consolidation [4].

Early crisis corpora supported supervised classification across multiple events, but discrepancies between independently constructed resources made direct comparisons difficult [5]. HumAID provides a large independently annotated multi-event dataset and evaluates classical, FastText and Transformer-based models under a common experimental setting [6].

CrisisBench, on the other hand, adopts a different approach by integrating eight existing datasets and mapping compatible labels to a shared benchmark task [4]. This consolidation provides broader event coverage under more consistent task definitions. However, this coordination requires decisions on which labels should be merged, which records should be retained, and which distinctions from the original datasets should be discarded. Consequently, the results reflect performance under the CrisisBench mapping scheme, rather than performance based on a universal disaster information classification system.

### B. Lightweight and Contextual Models

In this study, 'lightweight text classification' is an operational description of a class of supervised learning methods that can efficiently handle sparse text representations without requiring large pre-trained neural network architectures. For the purposes of this study, these methods are attractive because they can be trained on standard hardware, are reproducible under controlled conditions, and require relatively few computational resources for evaluation.

Resource-aware evaluation of conventional classifiers has also considered predictive performance together with computational requirements. A real-time mobile personal-cloud study [34] compared conventional classifiers using predictive accuracy, model-building time, system load and per-classification processing time. Although its task and hardware environment differ from those of the present CrisisBench study, it provides a methodological precedent for evaluating classifier choice as a performance–resource trade-off.

Sparse, high-dimensional text features are well suited to linear classifiers [8], [10], although TF-IDF and short word n-grams do not directly represent conversational history, broader event context or implicit semantic relationships.

The limitations of sparse lexical representations have encouraged contextual language models for crisis-related classification. Transformer architectures provide the foundation for such contextual representations [25]. CrisisBERT applied a Transformer-based architecture to crisis detection and crisis recognition [11], CrisisTransformers introduced domain-specific pre-training for crisis-related social-media text [12], and CrisisSense-LLM used instruction fine-tuning for multi-dimensional disaster text classification [13]. These approaches provide the contemporary contextual-model setting against which the present lightweight investigation is positioned. Differences in datasets, labels and experimental protocols mean that their reported scores are used as research context rather than direct numerical baselines.

Domain-specific AI has also been investigated beyond crisis NLP, including oceanic eddy identification [14], trajectory prediction [15], broader ocean-science AI frameworks [29], and deep-sea debris detection [30]. These studies provide broader examples of domain-specific model design rather than direct performance baselines for the present task.

### C. Research Gap

This study targets a specific empirical gap at the intersection of humanitarian text classification and resource-aware model evaluation. CrisisBench and HumAID provide benchmark evidence for classical and neural crisis classifiers [4], [6], while previous resource-aware research demonstrates the value of evaluating predictive performance together with computational requirements [34]. Building on these strands, the present study combines lightweight classifier and representation comparison with label-sensitive evaluation, directional error analysis, relative decision margins, measured computational requirements and domain-specific tropical-cyclone training within one controlled CrisisBench protocol.

This integrated analysis is important because aggregate benchmark scores alone cannot show whether performance is distributed evenly across a strongly imbalanced humanitarian label scheme, which label distinctions drive recurring errors, how prediction ambiguity relates to classification accuracy, or how restricting model scope changes the performance–resource trade-off.

Beyond classification, emergency information may also need to operate under disrupted connectivity. Opportunistic networking supports content dissemination under intermittent connections [18], while MANETs [19] and delay-tolerant networking [20] provide mechanisms for dynamic or store-and-forward communication. Related post-disaster communication research has also highlighted reliability and trust challenges in critical communications [21].

A-FC [32] provides a concrete example of a complementary emergency-communication layer. It is an activity-based delay-tolerant routing protocol designed for emergency communications in campus environments, in which store-and-forward mechanisms support information delivery when conventional connectivity is disrupted. In the research lineage considered here, lightweight semantic classification and A-FC-style communication address complementary stages of a wider disaster-information workflow: the classifier organises message content at the application layer, while disruption-tolerant mechanisms provide a potential means of transporting selected information. This relationship motivates a future direction for integrating semantic classification with disruption-tolerant information delivery.

## III. DATASET AND EXPERIMENTAL METHOD

### A. Dataset and Task

This study utilises the CrisisBench v1.0 English humanitarian classification dataset. CrisisBench integrates eight annotated crisis datasets, maps compatible source labels into shared schemes for informativeness and humanitarian classification, and provides predefined benchmark splits [4]. The study uses the consolidated English-language humanitarian classification files provided within CrisisBench v1.0 [4].

The selected task contains 87,455 English-language social-media records and is defined as a supervised single-label multi-class classification task. Given cleaned post text, the classifier predicts one of the 11 labels in the CrisisBench humanitarian classification scheme [4]. All 11 labels were retained without merging.

| Split | Number of records | Experimental role |
|---|---|---|
| Train | 61,164 | Model fitting for initial and Development-stage experiments |
| Development | 8,935 | Model comparison, tuning and class-weight experiments |
| Test | 17,356 | Held-out final evaluation after configuration locking |
| Total | 87,455 | Complete selected task |

Table 1: CrisisBench English Humanitarian Data Splits

The Test split was reserved until the configuration had been selected using Train and Development data. Only text_clean was used as predictive input; event and source were retained for descriptive and supplementary analyses and excluded from model features. The id field was removed because it was not required for modelling and may serve as an online identifier, while lang and lang_conf were excluded because the selected files had already been filtered to English. Public availability does not remove privacy and identifiability considerations in internet research [27]; row-level text was therefore handled as restricted research material in accordance with the project data-management arrangements.

All three processed splits contained the expected 11 labels and no missing or empty modelling text. No exact duplicate text was found within individual splits, and no exact cleaned-text overlap was found between Train, Development and Test.

### *B. Preprocessing and Feature Representation*

The same deterministic preprocessing function was applied to the Train, Development and Test text fields. The procedure was deliberately conservative: HTML entities were decoded; an initial RT marker was removed; URLs and user mentions were replaced with url and user; # characters were removed while retaining hashtag text; whitespace was normalised; and the text was converted to lowercase.

No stop-word filtering, stemming, lemmatisation or separate custom tokenisation step was applied. Punctuation, emojis and numerals were not aggressively removed.

Word-level TF-IDF [7] was used with unigram and unigram–bigram representations. The vectoriser used min_df=2, max_df=0.95, sublinear term frequency, L2 normalisation and float32 sparse features. During Development-stage experiments, each vectoriser was fitted only on Train and applied unchanged to Development. After configuration locking, a new vectoriser was fitted on the combined Train and Development text before application to Test.

### *C. Experimental Protocol and Model Selection*

This study employs a controlled supervised text-classification design based on the predefined CrisisBench Train, Development and Test splits. Train was used for initial fitting, while Development was used to compare classifier families, TF-IDF representations, regularisation settings and class weights. Macro-F1 served as the primary model-selection metric, while Accuracy, Weighted-F1 and per-label measures provided complementary evidence. Test was excluded from all model-selection decisions. Once a configuration had been selected and locked, Train and Development were combined to fit the final model, which was then evaluated on the held-out Test split.

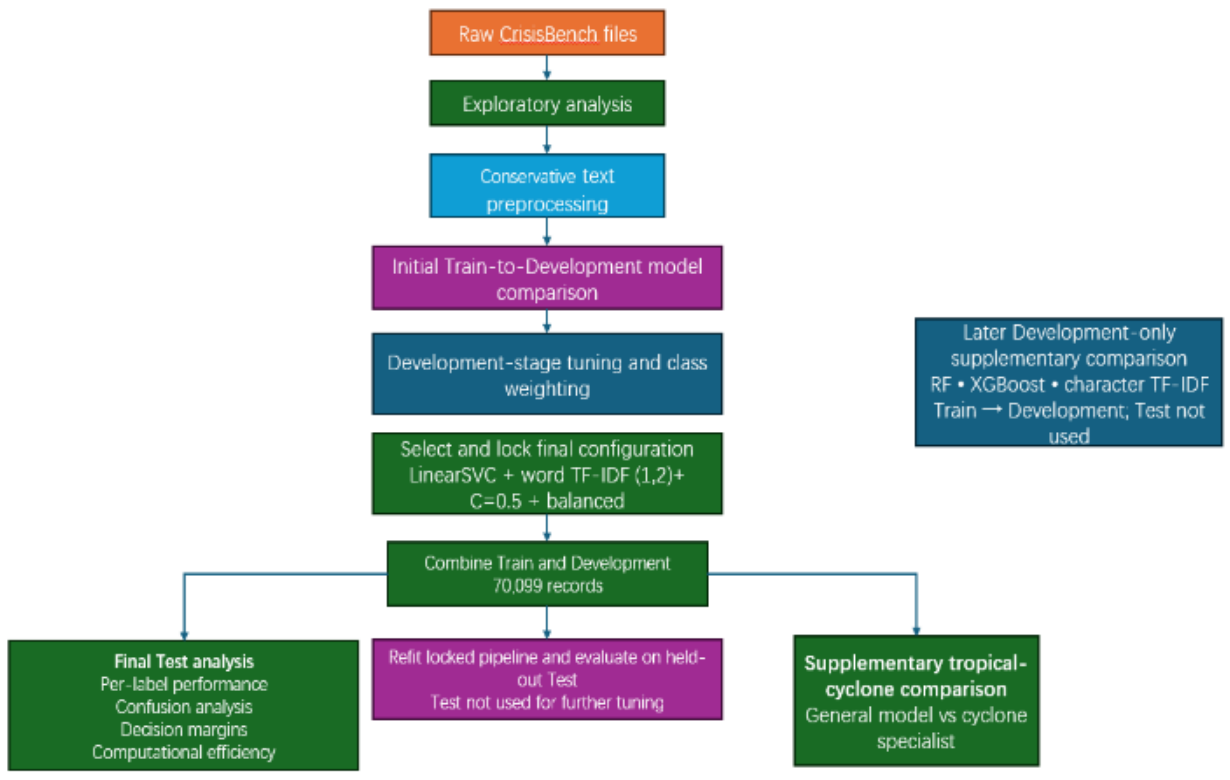

Figure 2: Experimental protocol for the CrisisBench humanitarian classification experiments

The methodological choices were linked to the specific challenges of the CrisisBench task. Label imbalance motivated the use of Macro-F1 and class-weight comparisons; short and lexically diverse posts motivated the comparison of word-unigram and unigram–bigram TF-IDF representations; overlap between humanitarian labels motivated per-label and confusion analysis; relative decision margins were analysed as an indicator of prediction ambiguity; and feature dimensionality, sparse-matrix memory use and recorded execution time were measured to examine computational requirements.

The experimental TF-IDF and classification pipelines were implemented using scikit-learn [26] and run locally on a Windows 11 laptop with an AMD Ryzen 9 7940HX CPU. Vectorisation, model fitting and prediction times were measured using Python's perf_counter; the resulting timings are treated as comparative measurements within this study rather than hardware-independent benchmarks.

The preliminary comparison included a majority-class DummyClassifier, Logistic Regression [10], an SGD hinge classifier [24] and Multinomial Naive Bayes [9]. With the exception of the majority baseline, all initial classifiers used word-unigram TF-IDF, with vectorisers fitted on Train and applied unchanged to Development. Class weighting was not used during this preliminary comparison.

Controlled Development-stage tuning subsequently evaluated Logistic Regression [10], LinearSVC [8] and Multinomial Naive Bayes [9]. Two word-level TF-IDF representations were evaluated: unigrams and combined unigrams–bigrams. Logistic Regression and LinearSVC were each evaluated with C=0.5, 1.0 and 2.0, both without class weighting and with class_weight="balanced". Multinomial Naive Bayes was evaluated with alpha=0.5 and 1.0 for both feature representations. Macro-F1 remained the primary selection criterion.

| Component | Values evaluated | Locked setting |
|---|---|---|
| Classifier | Logistic Regression, LinearSVC and Multinomial Naive Bayes | LinearSVC |
| Word n-grams | (1, 1) and (1, 2) | (1,2) |
| Linear-model C | 0.5, 1.0, 2.0 | 0.5 |
| Class weighting | None and balanced | balanced |
| Naive Bayes alpha | 0.5, 1.0 | Not applicable |
| Primary selection metric | Macro-F1 | Macro-F1 |
| Secondary measures | Accuracy and Weighted-F1 | Used for supporting comparison |

Table 2: Development-Stage Search And Locked Configurations

The design produced 28 Train-to-Development runs. Every vectoriser and classifier was fitted only on Train and evaluated on the unchanged Development split. Configurations were ranked primarily by Macro-F1, with Accuracy and Weighted-F1 used as supporting measures. The selected and locked configuration was LinearSVC with word unigram–bigram TF-IDF, C=0.5 and balanced class weighting.

After configuration locking, Train and Development were combined, producing 70,099 fitting records. A new TF-IDF vectoriser and LinearSVC were fitted using the locked settings. Test remained excluded from all configuration decisions.

### D. Evaluation Methods

Model evaluation used Accuracy, Macro-F1 and Weighted-F1, supported by per-label Precision, Recall and F1 [23]. Macro-F1 was the primary measure because each of the 11 labels contributed equally, reducing the influence of the frequent not_humanitarian class.

Confusion matrices were generated using counts and row-normalised values. Per-label statistics included support, precision, recall, F1 score and error rate, while high-frequency directional confusion pairs were used to identify recurring errors.

For the final LinearSVC model, the decision margin was calculated as the difference between the highest and second-highest decision scores. Test records were divided into five equally sized groups based on margin ranking. The margin was treated solely as a relative indicator of score separation and prediction ambiguity, rather than as a calibrated probability or operational confidence threshold.

Computational efficiency was assessed using feature dimensionality, sparse-matrix memory requirements, vectorisation time, model-fitting time and prediction time. For the supplementary Development comparison, training-pipeline time and serialised vectoriser–classifier size were also recorded.

### E. Tropical-Cyclone and Supplementary Comparisons

A supplementary domain-restricted experiment examined tropical-cyclone-related records. Records were selected using case-insensitive matching of the event field. The strict filter matched labels containing typhoon or hurricane, producing 20,103 records in total. The broad filter additionally matched cyclone, producing 21,944 records; this broad definition was used for specialist-model training and the primary general-versus-specialist comparison. The 13 source-specific event-label strings identified by the broad filter corresponded to nine named tropical cyclones rather than 13 independent storms. The named cyclone events were also cross-checked against NOAA Historical Hurricane Tracks [33].

The general model used the locked configuration fitted on the combined full Train and Development data. Its Test predictions were filtered to the broad and strict cyclone subsets. The specialist model used the same locked TF-IDF and LinearSVC settings but was fitted only on broad tropical-cyclone Train and Development records. It was evaluated on the broad cyclone Test subset, with its predictions additionally filtered to the strict subset as a sensitivity comparison. No separate strict specialist model was trained.

A separate supplementary experiment was conducted after completion of the main model-selection and held-out Test protocol. Train (61,164 records) was used for fitting and Development (8,935 records) for evaluation. Four fixed configurations were compared: the locked word-level LinearSVC reference, Random Forest [35], XGBoost [36], and LinearSVC with character-level TF-IDF. Test was neither loaded nor evaluated, so this experiment did not reopen the completed final-model selection procedure.

| Model | Representation | Main settings |
|---|---|---|
| LinearSVC reference | Word TF-IDF (1,2) | C=0.5; class_weight=balanced |
| Random Forest | Word TF-IDF (1,2) | 100 trees; max_features=sqrt; class_weight=balanced_subsample; n_jobs=-1 |
| XGBoost | Word TF-IDF (1,2) | 100 estimators; max_depth=6; learning_rate=0.1; subsample=0.8; colsample_bytree=0.8; balanced sample weights; tree_method=hist; device=cpu |
| LinearSVC | Character TF-IDF (3–5) | analyzer=char_wb; C=0.5; class_weight=balanced |

Table 3: SUPPLEMENTARY DEVELOPMENT-ONLY CONFIGURATIONS

The supplementary configurations were evaluated using Accuracy, Macro-F1 and Weighted-F1. Resource measurements included feature dimensionality, sparse Train-matrix size, vectorisation time, classifier fitting time, Development prediction time and serialised pipeline size. No additional hyperparameter search was performed.

## IV. RESULTS

### A. Development-Set Results

The Development split was used for model comparison and configuration selection, while the Test split remained excluded from this stage. Macro-F1 was the primary selection metric because performance across all 11 labels was of interest under the strongly uneven class distribution. Accuracy and Weighted-F1 were retained as complementary measures.

| Model | Accuracy | Macro-F1 | Weighted-F1 |
|---|---|---|---|
| Majority-class baseline | 0.5887 | 0.0674 | 0.4363 |
| Logistic Regression | 0.8263 | 0.5971 | 0.8128 |
| SGD hinge classifier | 0.8040 | 0.5369 | 0.7833 |
| Multinomial Naive Bayes | 0.7292 | 0.3757 | 0.6794 |

Table 4: Initial classifier performance on the CrisisBench Development split

The majority-class baseline achieved an Accuracy of 0.5887 but a Macro-F1 of only 0.0674, illustrating how the dominant not_humanitarian label could produce a moderate overall Accuracy without useful performance across the complete label scheme. Logistic Regression produced the strongest initial result, with an Accuracy of 0.8263 and a Macro-F1 of 0.5971. The SGD hinge classifier reached a Macro-F1 of 0.5369, while Multinomial Naive Bayes achieved 0.3757. Weighted-F1 exceeded Macro-F1 for all three text classifiers, indicating stronger performance on the more frequent labels.

The controlled tuning stage evaluated 28 Train-to-Development configurations across Logistic Regression, LinearSVC and Multinomial Naive Bayes. The experiments varied word n-gram representation, the regularisation parameter for the linear models, class weighting and Naive Bayes smoothing. Table 5 presents the configurations most relevant to the final selection; the complete set of runs was retained in the experimental outputs.

| Model | Word n-grams | Parameters | Class weighting | Accuracy | Macro-F1 | Weighted-F1 |
|---|---|---|---|---|---|---|
| LinearSVC | (1, 1) | C=0.5 | None | 0.8322 | 0.6176 | 0.8218 |
| LinearSVC | (1, 1) | C=0.5 | Balanced | 0.8156 | 0.6352 | 0.8217 |
| LinearSVC | (1, 2) | C=0.5 | None | 0.8360 | 0.6119 | 0.8253 |
| LinearSVC | (1, 2) | C=0.5 | Balanced | 0.8245 | 0.6413 | 0.8272 |
| Logistic Regression | (1, 1) | C=2.0 | None | 0.8327 | 0.6161 | 0.8219 |
| Multinomial Naive Bayes | (1, 1) | α = 0.5 | Not applicable | 0.7708 | 0.4503 | 0.7388 |

Table 5: Selected CrisisBench Development configurations for matched comparison and final model selection

LinearSVC accounted for all ten of the top-ranked configurations by Macro-F1. The best configuration, which combined unigram–bigram TF-IDF (C=0.5) with balanced class weights, achieved an Accuracy of 0.8245, a Macro-F1 of 0.6413 and a Weighted-F1 of 0.8272. The matched configurations show that neither adding bigrams nor applying balanced class weights improved every metric in every setting. In the relevant LinearSVC comparisons, balanced weighting increased Macro-F1 while reducing Accuracy relative to the corresponding unweighted configuration, whereas the effect of adding bigrams depended on the class-weight setting.

Compared with the strongest initial model, Logistic Regression, the selected LinearSVC increased Development Macro-F1 from 0.5971 to 0.6413, while Accuracy changed from 0.8263 to 0.8245. The unweighted unigram–bigram LinearSVC achieved the highest Development Accuracy of 0.8360, whereas the balanced configuration achieved the highest Macro-F1 of 0.6413 among the evaluated configurations. Because Macro-F1 had been defined in advance as the primary selection criterion, the balanced unigram–bigram LinearSVC was selected and locked before Test evaluation. No further tuning was performed.

## B. *Final CrisisBench Test and Per-Class Performance*

After Development-stage model selection, the chosen configuration was locked and no further parameter changes were made. Train and Development were then combined, providing 70,099 records for final model fitting. The resulting LinearSVC classifier was evaluated on the 17,356 held-out CrisisBench Test records. No Test result was used to reopen model selection or tuning. Table 6 compares this formal Test result with the Development result obtained during configuration selection.

| Metric | Development | Test | Test minus Development |
|---|---|---|---|
| Accuracy | 0.8245 | 0.8282 | +0.0037 |
| Macro-F1 | 0.6413 | 0.6578 | +0.0165 |
| Weighted-F1 | 0.8272 | 0.8315 | +0.0043 |

Table 6: Locked-model performance on the CrisisBench Development and held-out Test splits

The locked model achieved an Accuracy of 0.8282, a Macro-F1 of 0.6578 and a Weighted-F1 of 0.8315 on the CrisisBench Test split. These values were close to the corresponding Development results, indicating that the performance observed during model selection was broadly retained after configuration locking.

The differences between Development and Test should be interpreted descriptively rather than as an independent improvement, because the final model was refitted on the larger combined Train and Development set and evaluated on a different partition. Weighted-F1 nevertheless remained substantially higher than Macro-F1, indicating that performance was stronger on the more frequent labels than uniformly across the 11-label scheme.

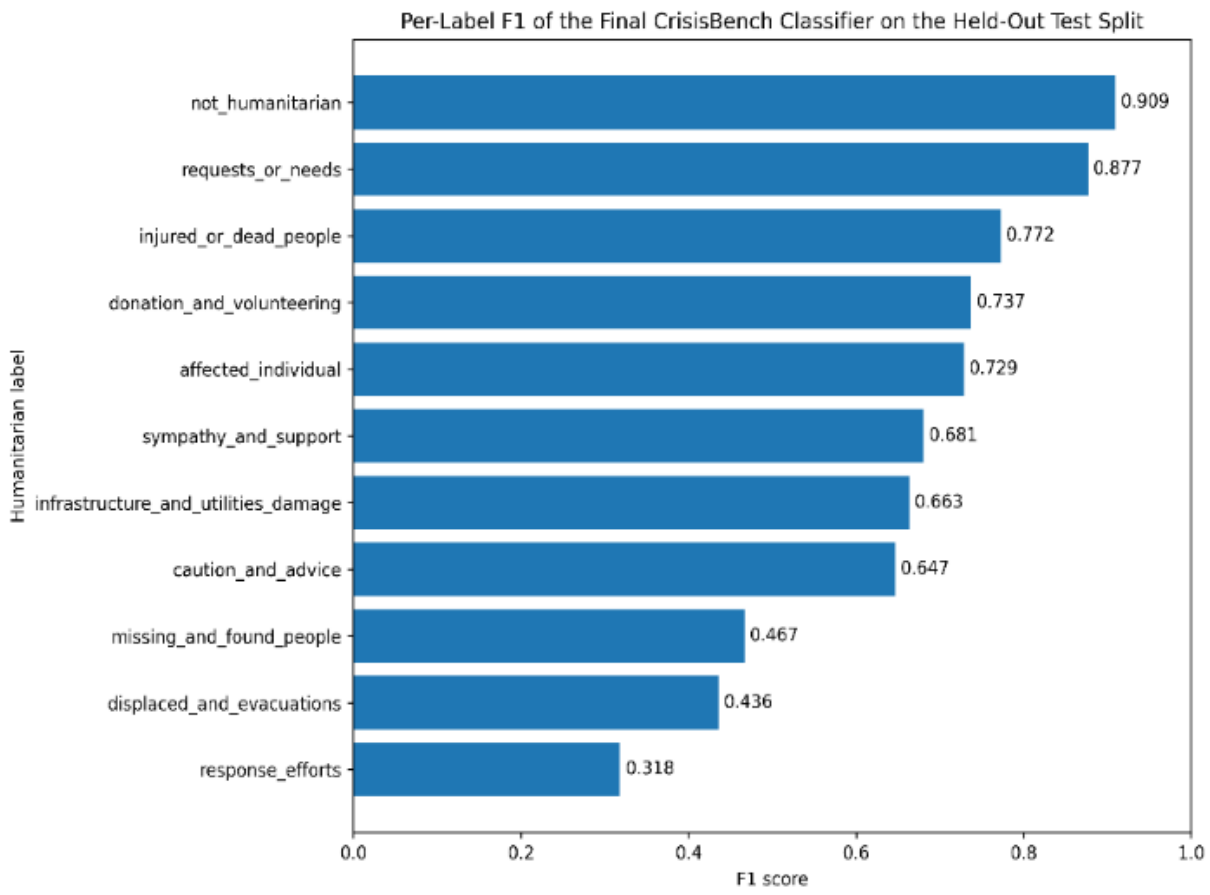

Figure 3: Per-label F1 scores of the final LinearSVC classifier on the CrisisBench held-out Test split

Figure 3 makes the uneven label-level performance visible. not_humanitarian achieved the highest F1 at 0.9095, followed by requests_or_needs at 0.8767 and injured_or_dead_people at 0.7724. In contrast, response_efforts achieved 0.3183, while displaced_and_evacuations and missing_and_found_people reached 0.4356 and 0.4667 respectively. The two latter labels had only 99 and 103 Test records, but support alone did not determine performance: injured_or_dead_people achieved comparatively strong performance with 561 records, whereas response_efforts remained weak with 221.

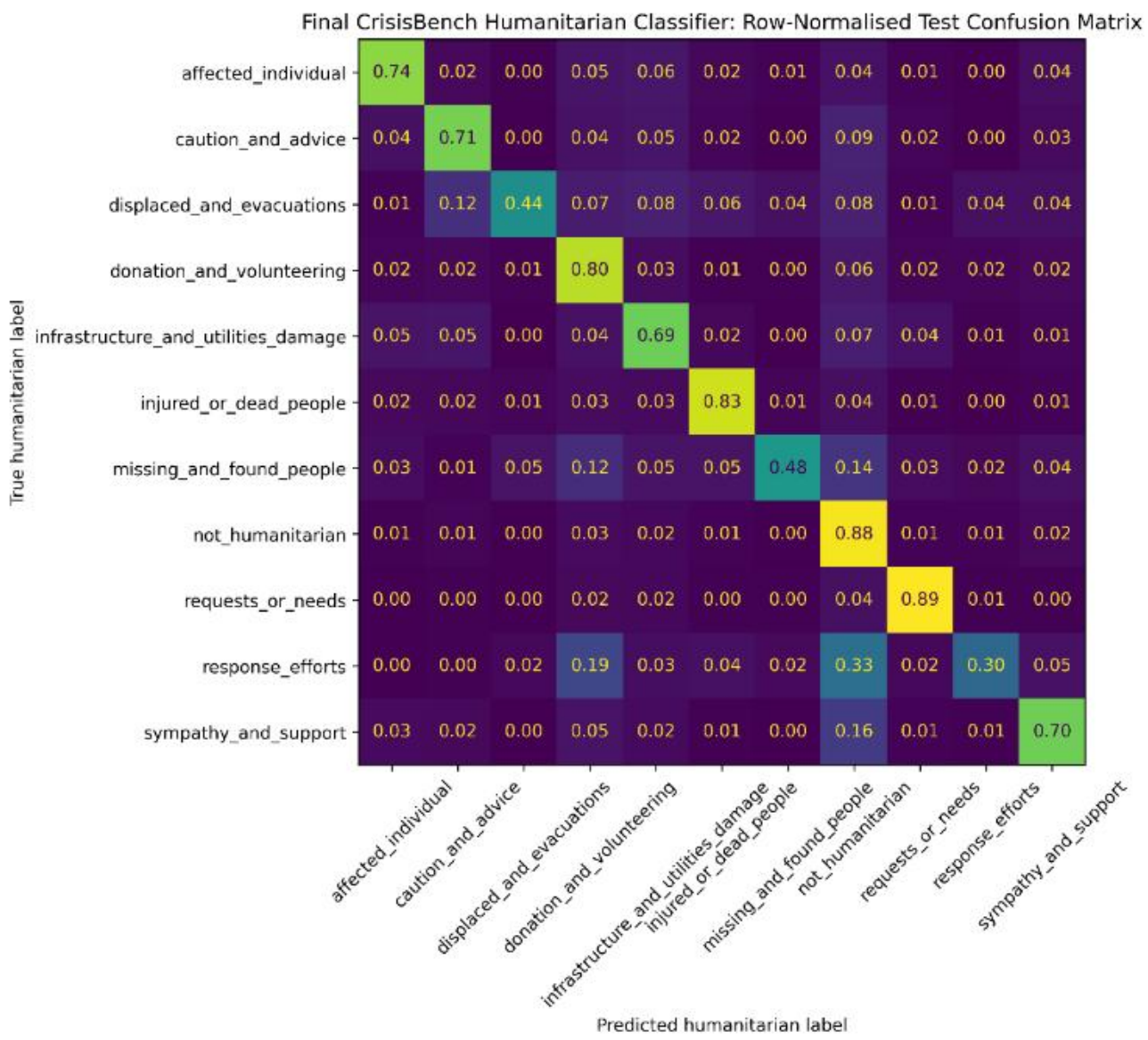

Figure 4: Row-normalised Test confusion matrix for the final CrisisBench humanitarian classifier.

The confusion matrix shows that errors were not distributed evenly across the humanitarian labels. Of the 221 true response_efforts records, 72 were predicted as not_humanitarian, corresponding to 32.58% of that class. Similarly, 162 sympathy_and_support records were assigned to not_humanitarian, accounting for more than half of the errors within that label.

The largest absolute error counts originated from the much larger not_humanitarian class, including 273 predictions as donation_and_volunteering, 242 as sympathy_and_support and 200 as infrastructure_and_utilities_damage. Each represented less than 3% of the 10,253 true not_humanitarian records, so their large absolute counts partly reflected the size of the majority class. Errors also occurred between humanitarian labels: for example, 53

infrastructure_and_utilities_damage records were predicted as caution_and_advice and 47 as affected_individual.

In a crisis-information setting, these patterns matter because potentially relevant response, support or damage-related messages may be absorbed into broader or non-humanitarian categories.

### C. *CrisisBench Test Decision Margins*

For each CrisisBench Test prediction, the LinearSVC decision margin was calculated as the difference between the highest and second-highest decision scores. It was analysed only as a relative indicator of prediction ambiguity and was not interpreted as a calibrated probability.

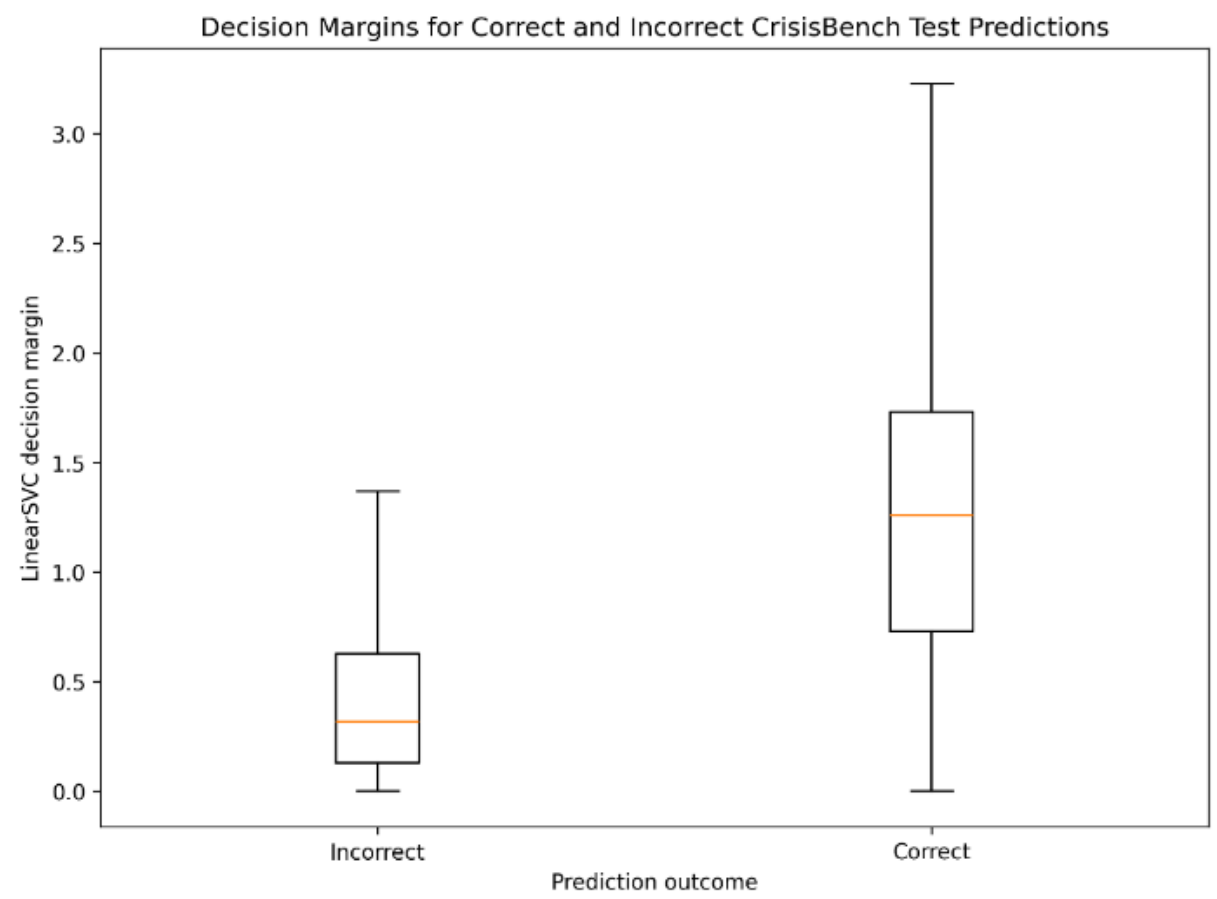

Figure 5: Decision margins for correct and incorrect humanitarian-label predictions on the CrisisBench Test split

Figure 5 shows that correct predictions generally had larger decision margins than incorrect predictions. The median margin was 1.2594 for correct predictions and 0.3167 for incorrect predictions. When Test records were divided into five equally sized groups by margin rank, Accuracy increased from 0.5036 in the lowest-margin quintile to 0.7663, 0.9162, 0.9660 and finally 0.9888 in the highest-margin quintile. Lower-margin predictions were therefore substantially more error-prone, although overlap between the distributions means that the margin cannot determine the correctness of an individual prediction. These results support the use of decision-margin ranking as a practical signal for prioritising uncertain predictions for human review. The margin remains a relative score-separation measure rather than a calibrated probability of correctness.

### D. *CrisisBench Tropical-Cyclone Model Comparison*

The general model and the cyclone specialist were evaluated on the same broad tropical-cyclone Test subset.

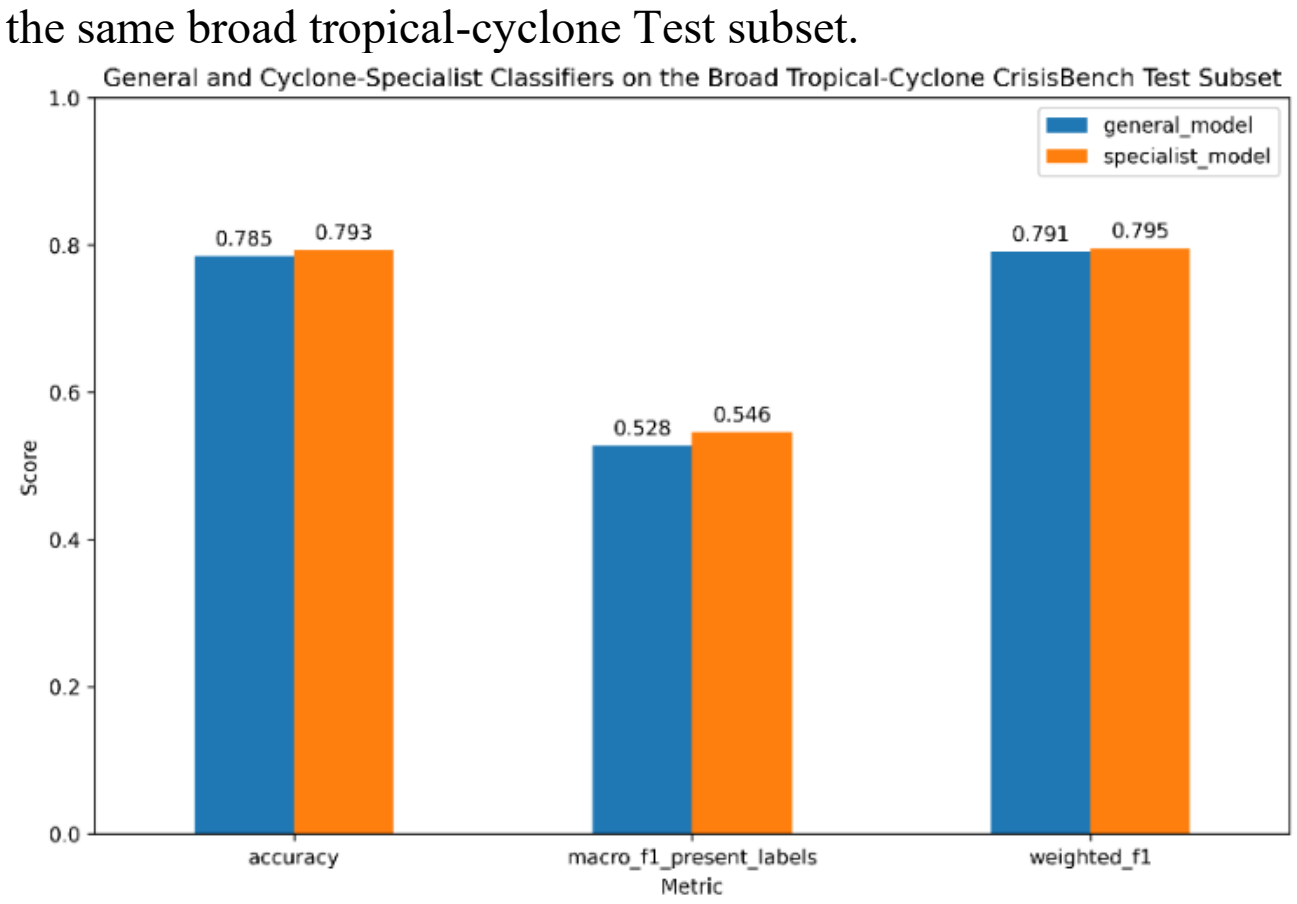

Figure 6: General and cyclone-specialist humanitarian classifiers on the broad tropical-cyclone CrisisBench Test subset.

On the broad tropical-cyclone Test subset, the specialist classifier achieved slightly higher scores across all three aggregate metrics. Accuracy increased from 0.7852 to 0.7933, Macro-F1 from 0.5276 to 0.5455 and Weighted-F1 from 0.7911 to 0.7955. The strict sensitivity check showed the same aggregate pattern, with Accuracy increasing from 0.7866 to 0.7976, Macro-F1 from 0.5105 to 0.5391 and Weighted-F1 from 0.7936 to 0.8005. The strict comparison used the same broad-trained specialist rather than a separately trained strict model.

Overall, the specialist model produced modestly higher aggregate scores on both cyclone subsets. However, the size of the improvement was limited, so the result supports a small in-domain benefit rather than the general superiority of the specialist model.

Aggregate improvements do not show whether the effect of specialist training was consistent across humanitarian labels. Fig. 7 therefore compares per-label F1 scores for the two classifiers on the broad tropical-cyclone Test subset.

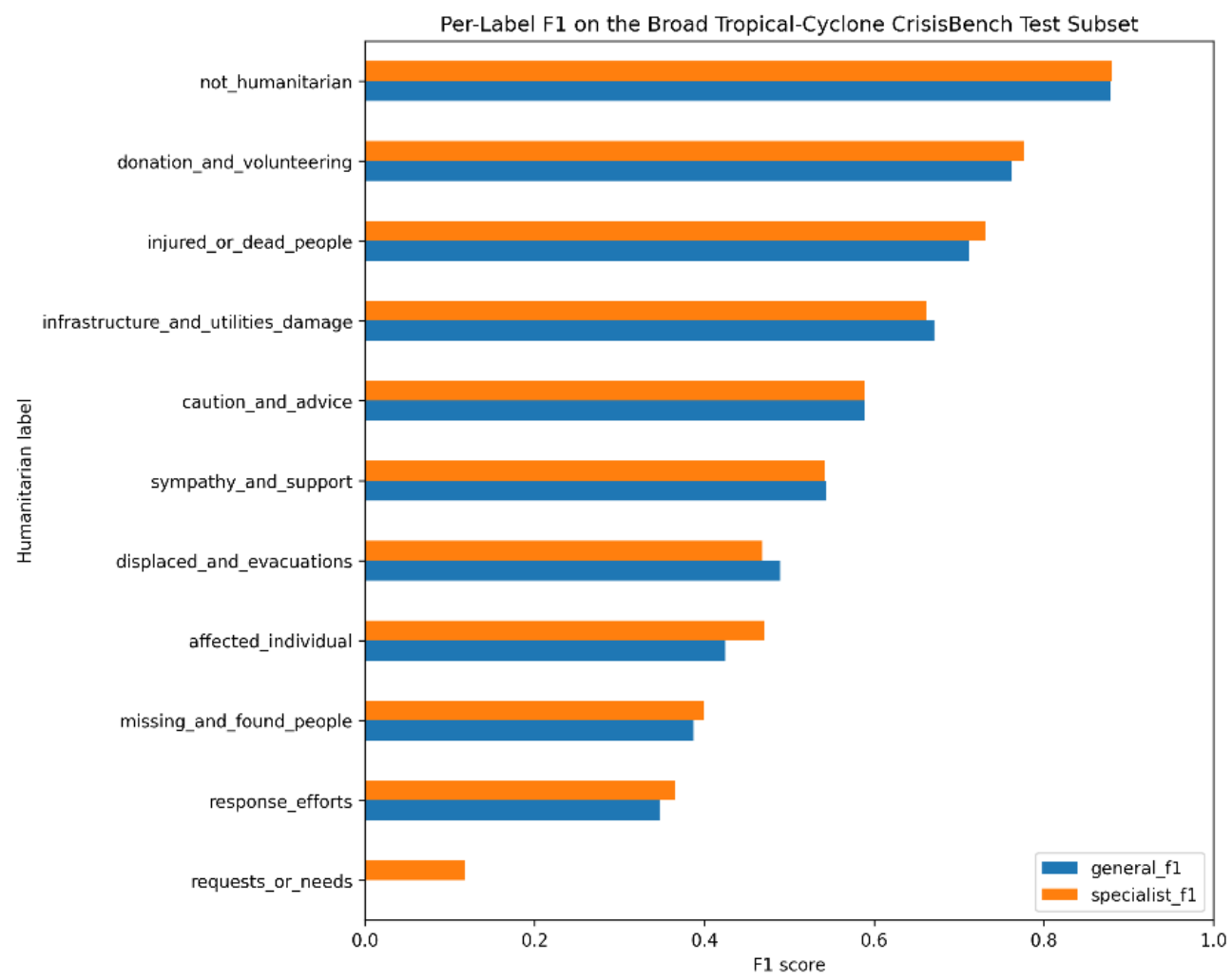

Figure 7: Per-label F1 comparison of the general and cyclone-specialist classifiers on the broad tropical-cyclone CrisisBench Test subset

The class-level comparison shows that performance differences between the general and specialist classifiers were not uniform across labels. The class-level comparison therefore characterises specialist training as a modest in-domain improvement whose effect varies across humanitarian labels.

### E. *Computational Efficiency*

Table 7 compares the computational measurements of the general and cyclone-specialist classifiers recorded within the same reproduced run. Using measurements from one run avoids mixing timings obtained under different execution conditions.

| Model | Training records | TF-IDF features | Training matrix | Vectorisation | Model training |
|---|---|---|---|---|---|
| General | 70,099 | 110,414 | 11.62 MB | 2.205 s | 1.876 s |
| Cyclone specialist | 17,586 | 30,168 | 2.58 MB | 0.508 s | 0.231 s |

Table 7: Computational comparison of the general and cyclone-specialist CrisisBench classifiers

The cyclone specialist used 17,586 training records compared with 70,099 for the general model. Its TF-IDF representation contained 30,168 features rather than 110,414, while the sparse training matrix occupied approximately 2.58 MB instead of 11.62 MB. Recorded vectorisation time decreased from 2.205 s to 0.508 s and fitting time from 1.876 s to 0.231 s. These reductions were consistent with the specialist model's smaller training set and narrower vocabulary. Because the general and specialist pipelines operated on datasets of different sizes, these timings describe the recorded workload for each model scope rather than normalised per-record speed.

Both classifiers completed prediction in less than 0.01 s on their respective Test sets, although prediction times should not be compared as direct speed ratios because the general model processed 17,356 records whereas the specialist processed 4,358. Overall, the specialist reduced feature, storage and fitting requirements considerably more than it changed predictive performance, while the general model retained broader multi-disaster coverage.

### *F. Supplementary Development Model and Resource Comparison*

A later Development-only experiment extended the comparison to Random Forest, XGBoost and a character-level LinearSVC representation, while retaining the selected word-level LinearSVC as the reference. Table 8 reports predictive performance together with training-pipeline time and serialised pipeline size. The Test split was not used in this comparison, so the previously completed final Test evaluation remained unchanged.

| **Configuration** | **Accuracy** | **Macro-F1** | **Weighted-F1** | **Training pipeline (s)** | **Serialised pipeline (MB)** |
|---|---|---|---|---|---|
| LinearSVC, word TF-IDF (1,2) | 0.8245 | 0.6413 | 0.8272 | 1.9965 | 10.15 |
| Random Forest, word TF-IDF (1,2) | 0.7455 | 0.4415 | 0.7063 | 18.3509 | 691.80 |
| XGBoost, word TF-IDF (1,2) | 0.6926 | 0.5445 | 0.7176 | 133.3794 | 4.58 |
| LinearSVC, character TF-IDF (3–5) | 0.8242 | 0.6546 | 0.8299 | 8.8156 | 16.14 |

Table 8: Additional CrisisBench Development model and resource comparison

The character-level LinearSVC achieved the highest Development Macro-F1 in this supplementary comparison at 0.6546, compared with 0.6413 for the word-level reference, while maintaining similar Accuracy. This higher Development Macro-F1 was accompanied by greater feature dimensionality, sparse-matrix size and recorded training-pipeline time. Under the fixed configurations evaluated here, Random Forest and XGBoost produced lower Macro-F1 than the word-level LinearSVC and required longer recorded training-pipeline times.

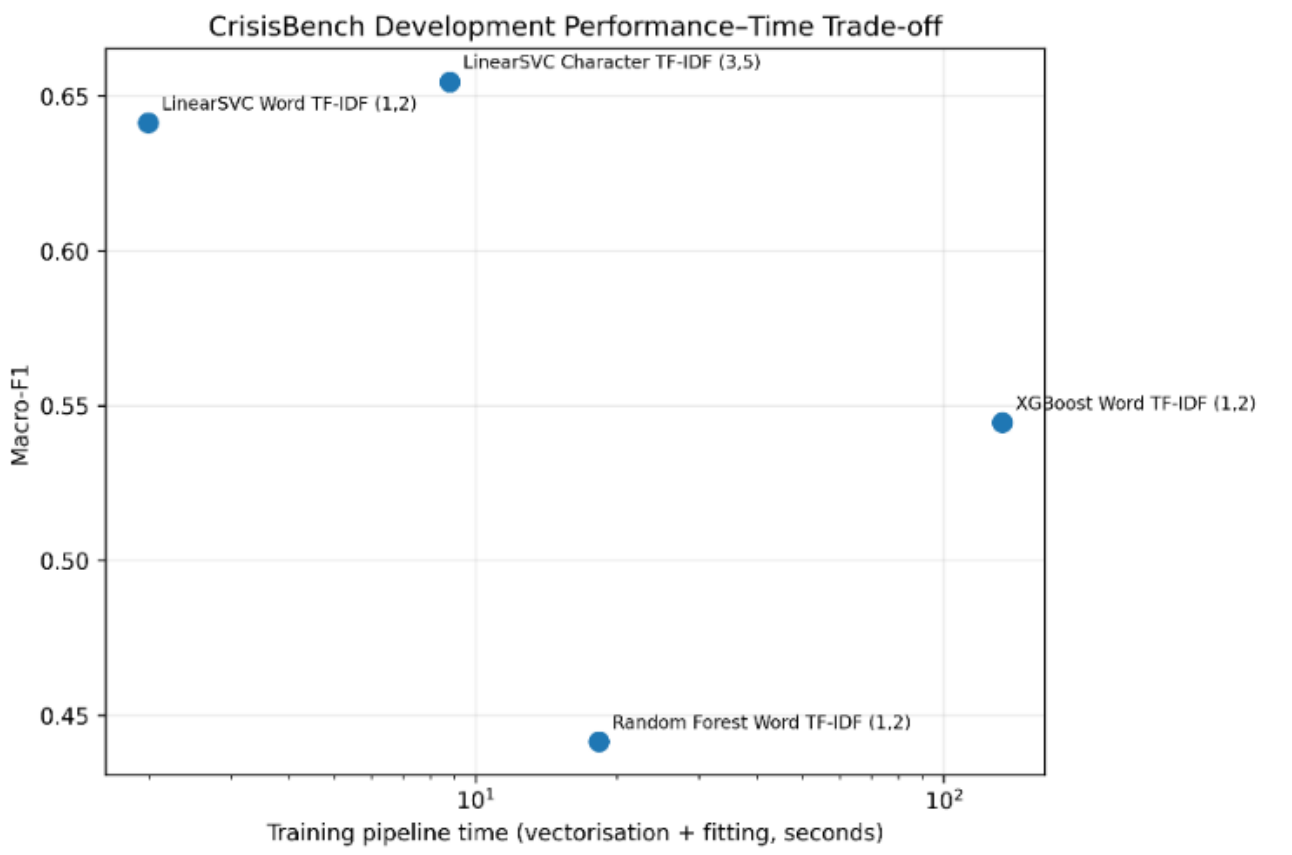

Figure 8: Development Macro-F1 versus recorded training-pipeline time for the supplementary CrisisBench model and representation comparison

## V. DISCUSSION AND LIMITATIONS

### *A. Effectiveness and Efficiency*

The final results from LinearSVC indicate that lightweight text classification can deliver useful overall performance on the CrisisBench 11-label humanitarian task. However, the Test Macro-F1 of 0.6578 was substantially lower than the Weighted-F1 of 0.8315, indicating that performance was not evenly distributed across the label scheme. The approach was effective as a broad classification baseline, but performance varied substantially across the humanitarian labels.

Model selection also illustrates why the evaluation criterion mattered. Logistic Regression achieved the strongest result in the initial comparison, but controlled Development-stage tuning favoured the balanced LinearSVC when Macro-F1 was prioritised. The selected configuration did not maximise Development Accuracy; instead, it provided the strongest balance across labels under the predefined selection criterion. This distinction is important for imbalanced humanitarian classification because optimising only the proportion of correct predictions can favour the dominant class [22].

The selected word-level LinearSVC also had relatively modest recorded resource requirements in the local experimental environment. In the supplementary Development comparison, character 3–5-gram TF-IDF achieved a higher Macro-F1 of 0.6546 than the word-level reference at 0.6413, but also required greater feature dimensionality, sparse-matrix size and recorded training-pipeline time. Under the fixed configurations evaluated, Random Forest and XGBoost did not exceed the word-level LinearSVC in Macro-F1 and required longer recorded training-pipeline times.

Resource-aware classifier selection has also been examined in earlier mobile-computing research, where conventional classifiers were evaluated using predictive accuracy together with model-building time, system load and per-classification processing time [34]. These absolute measurements are not directly comparable with the CrisisBench experiments because the tasks, hardware and timing scopes differ. Within the configurations evaluated here, the word-level balanced LinearSVC therefore provided a favourable performance–resource balance among the evaluated configurations, while the character-level representation offered a modest Development Macro-F1 gain at greater computational cost.

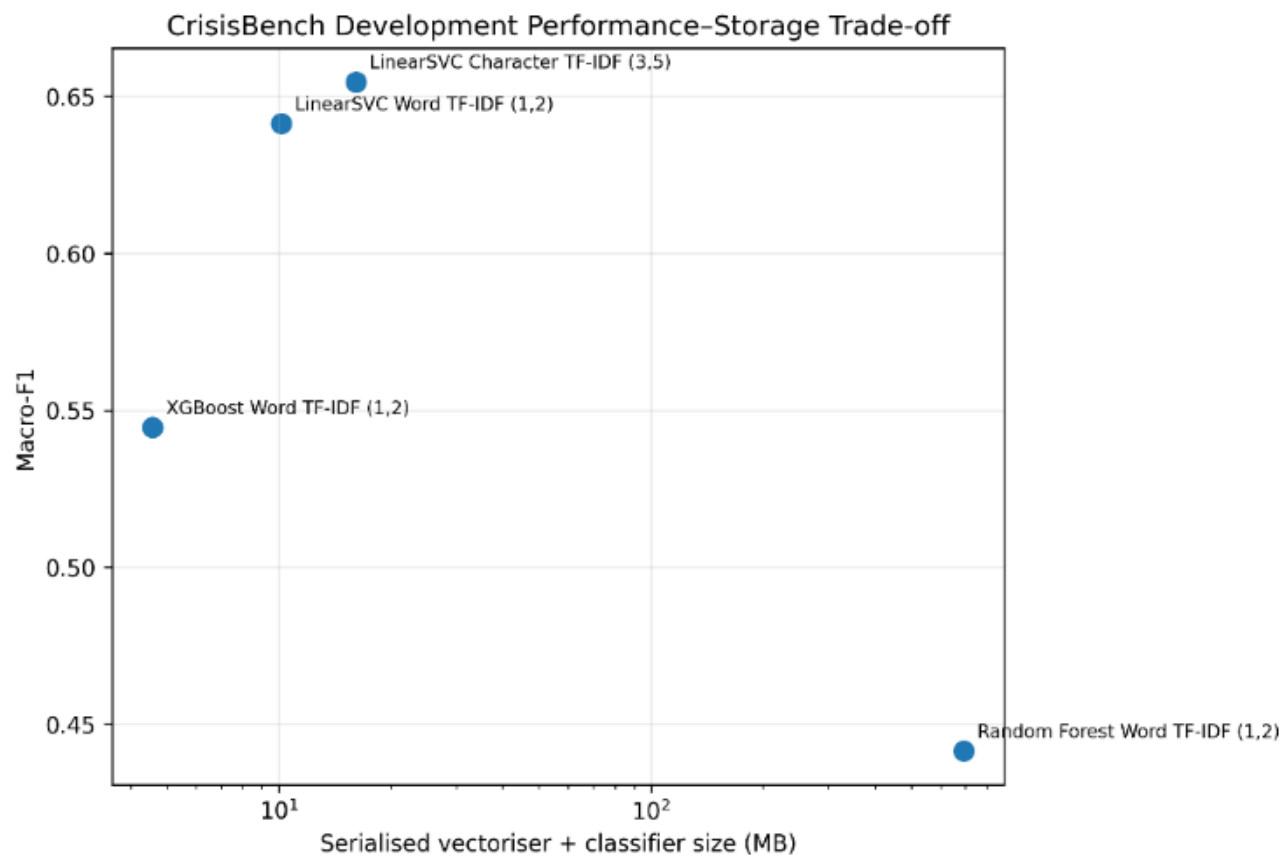

Figure 9: Development Macro-F1 versus serialised vectoriser–classifier size for the supplementary CrisisBench model and representation comparison

### *B. Class Imbalance, Error Behaviour and Human Review*

Class imbalance was associated with the unequal behaviour of the final CrisisBench classifier, but frequency alone did not explain the results. The dominant not_humanitarian label achieved an F1 of 0.9095, whereas response_efforts achieved only 0.3183. The two smallest labels, displaced_and_evacuations and missing_and_found_people, also performed comparatively weakly. However, injured_or_dead_people achieved substantially stronger performance despite having far fewer examples than the majority class. Label support was therefore relevant to performance, but it did not determine F1 outcomes on its own.

Semantic overlap and the single-label annotation scheme provide a plausible additional explanation. CrisisBench harmonises labels from several source datasets into one shared humanitarian task [4], while earlier crisis corpora also show that individual posts may contain several related forms of humanitarian information [5]. A post may describe infrastructure damage, affected individuals and a request for assistance within the same short message, even though the benchmark requires one final label. TF-IDF can represent recurring words and short phrases, but it cannot recover missing contextual information or determine which plausible interpretation should dominate. The later character-level comparison also showed that representation choice can affect aggregate Macro-F1, although the experiment did not determine which specific labels or linguistic phenomena were responsible for the improvement.

Balanced class weighting improved Development Macro-F1 for the selected LinearSVC, but it did not remove these semantic difficulties. Weighting changes the relative influence of training examples; it does not create additional minority-class evidence or resolve ambiguity in the annotation scheme.

The decision-margin analysis adds a further dimension to this label-sensitive evaluation by linking score separation with empirical error behaviour. Predictions in the lowest-margin quintile were substantially less accurate than those in the highest-margin quintile, indicating that margin ranking can help identify outputs that warrant closer human inspection. This provides a practical basis for prioritising uncertain predictions within a human-in-the-loop workflow, consistent with the broader principle demonstrated by AIDR [3]. The LinearSVC margin remains a relative score-separation measure rather than a calibrated probability, and an operational review threshold would require separate validation.

### C. General and Tropical-Cyclone Models

The tropical-cyclone specialist achieved modestly higher aggregate scores than the general classifier on the same in-domain Test subset. Broad-subset Macro-F1 increased from 0.5276 to 0.5455, and the strict sensitivity comparison showed the same direction. The experiment therefore shows a small but consistent aggregate in-domain benefit from specialist training within the two selected cyclone definitions.

The changes in computational requirements were more pronounced than the changes in predictive scores. The specialist used 17,586 training records and 30,168 TF-IDF features, compared with 70,099 records and 110,414 features for the general model. Its sparse training representation and fitting requirements were correspondingly smaller. These lower computational requirements were consistent with the specialist being trained on a narrower dataset and vocabulary, demonstrating a concrete performance-scope-resource trade-off within the same modelling configuration.

The effect of specialist training was nevertheless label-dependent. Some humanitarian labels improved while others did not, and narrowing the model to tropical-cyclone data also reduced its coverage of the broader mixture of disaster types represented by the general classifier. Domain-restricted training is therefore most relevant when incoming information is already known to concern tropical cyclones. This result is conceptually consistent with the wider motivation for domain-adapted crisis models [11]–[13]: restricting or adapting a model towards a particular domain can alter how task-relevant language is represented. The present experiment investigates this question using a much simpler TF-IDF and linear-classifier setting.

Overall, the tropical-cyclone experiment exposes a clear performance–scope–resource trade-off: specialist training produced modestly higher aggregate in-domain performance and substantially lower feature and fitting requirements, while the general classifier retained broader multi-disaster coverage. The current record-level Test design does not establish generalisation to tropical cyclones entirely absent from model development, which remains an important direction for future evaluation.

### D. Limitations, Ethics and Scope of Interpretation

Several limitations restrict the scope of the findings. First, the CrisisBench evaluation uses held-out records rather than a strict held-out-event protocol. Event groups represented in the Test split also appear elsewhere in the benchmark, so the results demonstrate generalisation to unseen records but not to disasters completely absent during model development. CrisisBench also harmonises several source datasets into a shared single-label scheme [4]. This improves comparability but can simplify messages containing more than one plausible humanitarian meaning.

Second, the modelling conclusions apply only to the methods and configurations actually evaluated. The main model-selection stage used word-level TF-IDF and lightweight supervised classifiers, while the later Development-only comparison added fixed Random Forest, XGBoost and character-level LinearSVC configurations without reopening the formal Test evaluation. These additional configurations were not subjected to extensive hyperparameter optimisation and should therefore not be interpreted as definitive rankings of the corresponding model families. No Transformer or large language model, external-dataset validation, probability calibration or statistical significance testing was included. The recorded feature, sparse-memory, timing and serialised-pipeline measurements are specific to the local experimental environment; electrical power consumption, energy use and peak process memory were not measured.

The use of social-media-derived text also creates privacy and research-ethics constraints. Public availability does not remove the need to consider how potentially identifiable data are acquired, processed, stored and reported [27]. Numeric identifiers were excluded from the processed files, URLs and user mentions were replaced, and raw or processed row-level text and predictions remained restricted in accordance with the project data-management arrangements. These measures reduce direct identifier exposure but do not guarantee complete anonymisation because text may still contain names, locations or descriptions of personal circumstances. For this reason, publicly presented examples should remain anonymised or paraphrased.

Finally, humanitarian classification is only one component of a potential disaster-information workflow. The model does not establish whether a post is true, current, geographically accurate or operationally actionable, nor does it determine rescue priority or resource allocation. Related post-disaster communication research also highlights reliability and trust as separate challenges in the delivery of critical information [21]. Any operational system would require additional processes for authorised data acquisition, verification, geographic interpretation, prioritisation and human review. The completed work should therefore be interpreted as an offline application-layer classification study rather than an end-to-end emergency-response system.

## VI. CONCLUSION AND FUTURE WORK

### A. Conclusion

This study investigated lightweight supervised machine learning for the CrisisBench v1.0 English humanitarian classification task under the combined challenges of limited textual context, semantic label overlap and severe class imbalance. A controlled Train–Development–Test protocol was used to compare lightweight classifiers and TF-IDF representations, with Macro-F1 as the primary model-selection criterion. The selected configuration was a balanced LinearSVC using unigram–bigram TF-IDF features with C=0.5, locked before the formal Test evaluation.

On the held-out CrisisBench Test split, the final classifier achieved an Accuracy of 0.8282, a Macro-F1 of 0.6578 and a Weighted-F1 of 0.8315. The difference between Macro-F1 and Weighted-F1, together with the per-label and confusion analyses, showed that strong aggregate performance did not imply uniform performance across the 11-label scheme. Decision-margin analysis further showed a strong empirical relationship between score separation and prediction errors, while the tropical-cyclone specialist produced modestly higher in-domain aggregate performance with substantially lower feature and fitting requirements.

The empirical contribution of the study is therefore a controlled, label-sensitive and resource-aware characterisation of lightweight humanitarian text classification. By combining model and representation comparison with class-level errors, decision margins, computational measurements and domain-specific training, the analysis identifies not only which lightweight configuration performs effectively, but also where classification remains difficult and how predictive performance changes with model scope and computational requirements.

## B. *Future Work*

Future work should first examine whether the findings generalise beyond the current CrisisBench benchmark and its historical source data. CrisisBench integrates crisis datasets collected in earlier periods, so changes in social-media language, platform conventions and disaster communication may limit how closely the benchmark represents current streams [4]–[6]. Evaluation on more recent datasets, together with event-level hold-out experiments in which complete disasters are excluded from model development, would provide stronger evidence of temporal and cross-event generalisation. Compact Transformer models or other crisis-specific language models [11]–[13] could also be compared under the same predictive and computational framework to determine when additional model complexity provides sufficient benefit to justify its resource requirements.

The observed relationship between decision margin and prediction errors also motivates further uncertainty analysis. Future studies could investigate calibrated probability estimates and validate review thresholds on separate data, allowing uncertainty information to be evaluated more systematically for human-review prioritisation [28].

A longer-term direction is to connect semantic classification with wider emergency-information delivery mechanisms. Resource-constrained deployment could investigate adaptive caching and emergency-communication frameworks [16], [17], opportunistic content dissemination [18], mobile ad hoc and delay-tolerant networking [19], [20], and energy-aware edge/fog resource management [31]. A-FC [32] provides a particularly relevant example of a complementary delay-tolerant communication layer that could support the transport of selected information when conventional connectivity is disrupted. Such integration would allow future work to study semantic classification and information delivery as complementary stages of a wider disaster-information workflow, with separate evaluation of predictive performance, delivery reliability, delay, storage, bandwidth and energy requirements.